\documentclass{qam-l}
\usepackage{amsmath, amsthm, graphics,
mathrsfs,latexsym,amscd,graphicx,amssymb, bm, upgreek}
\newcommand{\be}{\begin{equation}}
\newcommand{\ee}{\end{equation}}
\newcommand{\ba}{\begin{eqnarray}}
\newcommand{\ea}{\end{eqnarray}}

\newcommand{\pa}{\partial}
\newcommand{\f}{\frac}

\newtheorem{theorem}{Theorem}[section]

\theoremstyle{definition}

\theoremstyle{remark}
\newtheorem{remark}[theorem]{Remark}

\numberwithin{equation}{section}

\begin{document}

\title{ A new two-component system modelling shallow-water waves}
\author{Delia Ionescu-Kruse}
\address{Simion Stoilow Institute of Mathematics  of the Romanian Academy, Research Unit No. 6,\\
P.O. Box 1-764, 014700 Bucharest, Romania}
\email{Delia.Ionescu@imar.ro}





\date{}
\begin{abstract}

For  propagation of surface shallow-water
waves on irrotational flows,
we derive a new two-component system. The system is obtained by  a variational approach in the Lagrangian formalism.
The system has a non-canonical Hamiltonian formulation.
We also find its exact solitary-wave solutions.



\end{abstract}

\maketitle

\section{Introduction}

In this paper we obtain the following system of nonlinear partial differential equations
\begin{equation}
\left\{\begin{array}{ll}
u_t+3uu_x + HH_x=\left[H^2\left(uu_{xx}+u_{xt}-\f{u^2_x}{2}\right)\right]_x\\
\\
 H_t+(Hu)_x=0,
\end{array}
\right. \label{new}\end{equation}
with $x\in\mathbf{R}$, $t\in\mathbf{R}$,
$u(x,t)\in \mathbf{R}$, $H(x,t)\in \mathbf{R}$. \\
We start from a general dimensionless version of  the two-dimensional irrotational water-wave problem
with a free surface and a flat bottom. We focus on the motion of shallow-water waves, waves whose length is still large compared with the depth of the water in which they propagate.
In this shallow-water regime,  many two-component systems  have already been derived and studied.
One of them is the well-known Green-Nagdhi  system \cite{green&naghdi}
\begin{equation}
\left\{\begin{array}{ll}
u_t+uu_x+ HH_x=\f 1{3H}\left[H^3(uu_{xx}+u_{xt}-u^2_x)\right]_x\\
\\
 H_t+(Hu)_x=0,
\end{array}
\right. \label{gn}
\end{equation}
which models shallow-water waves whose amplitude (in the dimensionless version, the amplitude parameter, that is, the ratio of the wave amplitude to the depth of the water) is not necessarily small.
$u(x,t)$ represents the   horizontal velocity or the depth-averaged\footnote{The depth-averaged value of a quantity $q(x,z,t)$ is defined by $\bar{q}(x,t):=\f 1{H(x,t)}\int_0^{H(x,t)}q(x,z,t)dz$.}  horizontal velocity and $H(x,t)$ is the free upper surface.
The Green-Naghdi equations are mathematically well-posed in the sense that they admit solutions over the relevant time scale for any initial
data that are reasonably smooth (see \cite{li3}, \cite{alvarez}). The solution of the Green-Naghdi  equations provides a good approximation of the solution of the full water-wave problem  (see \cite{li3}, \cite{lannes1}).
The Green-Naghdi equations have nice structural
properties that facilitate the derivation of simplified model
equations in the shallow water regime. For example, the
celebrated  Korteweg-de Vries, Benjamin-Bona-Mahoney, Camassa-Holm and Degasperis-Procesi
equations arise as approximations to the Green-Naghdi equations
cf. the discussion in \cite{c&l}.

Actually, Green and Naghdi considered in  \cite{green&naghdi}    the three-dimensional water-wave problem with a free surface and a variable bottom, and no
assumption
of
an irrotational
flow
was made
a priori.
The equations were   derived  by
imposing the condition that the horizontal velocity is independent of   the vertical coordinate $z$, the condition that the vertical
velocity has only a linear dependence on  $z$ and  by using the mass conservation equation and the energy equation in integral form plus invariance under
rigid-body translation.
For one horizontal $x$-coordinate  and for a flat bottom, the equations have the form (\ref{gn}).
In the two-dimensional case (only one horizontal dimension) and for a domain with a flat bottom, the
system (\ref{gn}) was  originally derived in 1953 by Serre \cite{Serre}, and independently rediscovered  by  Su and Gardner
\cite{su&gardner} in 1969.
 Serre (\cite{Serre}, Sect. V.)  integrated the Euler equations over $z$ on the interval $[0,H(x,t)]$    and    made the assumption that the
  horizontal component of fluid velocity is  equal to its depth-averaged value.
Su and Gardner
\cite{su&gardner} obtained the system (\ref{gn})  by depth-averaging the  two-dimensional irrotational  water-wave problem and by using a  long-wave
asymptotic expansion. In the literature, the equations (\ref{gn}) are sometimes referred
to as the Serre equations, or the Su-Gardner equations but usually
they are called the Green-Naghdi equations.
Very recently, Ionescu-Kruse \cite{io2012'} obtained, by a variational approach in the Lagrangian formalism, the system (\ref{gn}) for the propagation of arbitrary
amplitude shallow-water waves on two-dimensional irrotational flows.

\noindent In Section 3 of the present paper, we  derive, by the same approach as in \cite{io2012'}, the  system (\ref{new}).
  We are in the shallow-water regime and we consider   surface waves of arbitrary amplitude.
   We are looking for a higher-order correction to the
  classical shallow-water equations (\ref{sw2}).
 The second equation of the system  (\ref{sw2}) is a transport equation,
 the free surface is advected or Lie transported (in the geometry literature), by the fluid
flow.
 In the system (\ref{new}), we keep  this equation as it is.   We obtain the first equation of the system (\ref{new})  by calculating the
critical points of an action functional in the space of paths with fixed endpoints,
within the Lagrangian formalism. We arrive at this action functional as follows. Within the Eulerian formalism, we consider the Lagrangian function integrated over time  in the action functional
to have the traditional form, that is, the kinetic energy minus the potential energy.
According to   a velocity field with a  horizontal component (\ref{13'}) independent of   the vertical coordinate $z$  and a  vertical component  (\ref{33})
 having only a linear dependence on  $z$,
  we take  for the kinetic energy at the free surface of the water the expression
 (\ref{energy})  and  for the potential energy calculated with respect to the
undisturbed water level the expression (\ref{21}).
Then,  we   transport the Lagrangian function (\ref{lagreuler}) from the Eulerian picture to
 the tangent bundle which represents the  velocity phase space in the Lagrangian formalism, this transport being made taking
 into account
 the second equation of
the system (\ref{new}) too. Thus, we get the Lagrangian function (\ref{lagr}).
We point out that the Lagrangian (\ref{lagreuler}) as well as (\ref{lagr}) are not metrics; the pursuit of an 
advanced geometrical approach is not necessarily dependent upon the existence of a metric, as illustrated in the recent papers 
\cite{escher} and \cite{escher-kolev} too.\\
The type of considerations made in the present paper proved also very useful (in similar contexts) to qualitative studies of some model 
equations. For example, in the derivation of criteria for global existence and blow-up of solutions as well as in studies of the propagation
speed for some model equations for shallow water waves, see e.g. the papers \cite{const2000}, \cite{c-e}, \cite{gui-liu}, \cite{const2005},
\cite{henry}.

The Green-Naghdi equations (\ref{gn}) have the following  Hamiltonian  formulation  (see \cite{holm}, \cite{const97})
\begin{equation}
\left(\begin{array}{c}
m_t\\
\\
 H_t\end{array}\right)=-\left(\begin{array}{cc}
\pa_x m+m\pa_x & H\pa_x\\
\\
 \pa_x H & 0\end{array}\right)\left(\begin{array}{c}
\f{\delta \mathcal{H}_{GN}}{\delta m}\\
\\
 \f{\delta \mathcal{H}_{GN}}{\delta H}\end{array}\right),
\label{hamgn}\end{equation}
where $ \mathcal{H}_{GN}$ is the total  energy (kinetic plus potential) given by
\be\f 1{2}\int_{-\infty}^\infty\left(
 H u^2+\f 1{3}H^3u^2_x+(H-1)^2\right)dx,\ee
and $m$ is the momentum density defined by
\be
m:=\f{\delta \mathcal{H}_{GN}}{\delta u}=Hu-\f 1{3}\left(H^3u_x\right)_x,
 \ee
$\f{\delta \mathcal{H}_{GN}}{\delta u}$, $\f{\delta \mathcal{H}_{GN}}{\delta m}$ and $\f{\delta \mathcal{H}_{GN}}{\delta H}$ being the variational derivatives
 of $\mathcal{H}_{GN}$ with respect to $u$, $m$ and $H$, respectively.

\noindent In Section 4 of the present paper, we show that the the  system (\ref{new}) has the Hamiltonian  formulation (\ref{hamgn}), with
a different total energy  $\mathcal{H}_{N}$ given  by (\ref{hn}) and a different momentum density $m$ given by (\ref{m}).

The solitary-wave solution of the Green-Naghdi equations (\ref{gn}) has the form (see \cite{Serre}, pag. 863-864 and \cite{su&gardner}, pag 539)
 \ba \begin{array} {ll} H(x,t)=1+(c^2-1)\,
\textrm{sech}^2\left[\f{\sqrt{3}}{2}\f{\sqrt{c^2-1}}{c}(x-ct)\right]\\
\\
u(x,t)=c\left(1-\f 1{H(x,t)}\right),\end{array} \label{solgn} \ea
with $c$ the  speed of the traveling wave.
 These waves exist for all $c$ such that the following condition:
\be
c^2>1 \label{cgn}
\ee
is satisfied. In \cite{li1}, \cite{li2},   the eigenvalue problem obtained from linearizing the
equations about solitary-wave solutions is  investigated and  it is established that
small-amplitude solitary-wave solutions of the Green-Naghdi
equations are linearly stable.

\noindent In Section 5 of the present paper, we find the  solitary-wave solution of the system (\ref{new}).  Its expression (\ref{sol1})-(\ref{sol2})
is different from (\ref{solgn}). The speed $c$ of the traveling wave has to satisfy the condition (\ref{36}), that is, the condition (\ref{cgn}).

 \section{Preliminaries}
We recall the classical water-wave problem for gravity waves propagating at the free surface of a two-dimensional  inviscid incompressible fluid.
The fluid occupies the domain:
\be
-\infty<x<\infty,\quad 0\leq z \leq h_0+\eta(x,t),
\ee
where the constant $h_0>0$ is the undisturbed depth of the water and $\eta(x,t)$ is the displacement of the free surface from the undisturbed state.
$(x,z)$ are the Cartesian coordinates,
 the $x$-axis being in the direction of wave propagation
and the $z$-axis pointing vertically upwards.
The governing equations are   Euler's equations and the continuity equation  with appropriate surface and bottom boundary conditions (see, for example,
\cite{const-carte}):
\begin{equation}
\begin{array}{cc}
u_t+uu_x+vu_z=- p_x&\\  v_t+uv_x+vv_z=- p_z-g&\\
 u_x+v_z=0&\\
 v=\eta_t+u\eta_x & \textrm{ on }\,
z=h_0+\eta(x,t)\\
 p=p_0&
\textrm{ on }\,
z=h_0+\eta(x,t)\\
 v=0 &
\textrm { on } \, z=0.
 \end{array}
\label{e+bc}
\ee
Here $(u(x,z,t), v(x,z,t))$ is the velocity field of the
water - no motion takes place in the $y$-direction, $p(x,z,t)$
denotes the pressure, $p_0$ being the constant atmospheric pressure and $g$ is the acceleration due  to gravity. We set the constant  density $\rho=1$.\\
The water flow is assumed to be irrotational, that is,  in addition to the system (\ref{e+bc}) we also have the equation
\be
u_z-v_x=0. \label{ic}
\ee
We introduce the following dimenionless variables (see, for example, \cite{johnson-carte}):
\be
\begin{array}{c}
\bar{x}=\f{x}{\lambda} ,  \quad \bar{z} = \f  {z}{h_0}, \quad \bar{t}=\f {\sqrt{gh_0}}{\lambda}t, \quad \bar{\eta}= \f{ \eta}{a},\\
\cr
  \bar{u}=  \f 1 {\sqrt{gh_0}}u,
\quad \bar{v}= \f 1{h_0}\f{\lambda}{\sqrt{gh_0}}v,
\\
\cr
\bar{p}= \f 1{gh_0}[p-p_0-g (h_0-z)],
\end{array} \label{nondim}\end{equation}
where $a$  represents a measure of the amplitude of the waves and $\lambda$ the typical
wavelength for the considered waves. The dimensionless variables considered above are good choices for
showing the  magnitude of the different terms that appear in the equations.
Substituting (\ref{nondim}) in the system (\ref{e+bc})-(\ref{ic}),
 one finds that the equations of motion depend upon  two parameters $\epsilon$ and $\delta$  defined as follows:
 \be
 \epsilon:=\f a{h_0}, \quad \delta:=\f {h_0}{\lambda}.
 \ee
The amplitude parameter $\epsilon$ is
associated with the nonlinearity of the wave, and the long-wave
parameter $\delta$ is associated with the dispersion of the wave.
Omitting the bars for the sake of clarity, the dimensionless form of the system (\ref{e+bc})-(\ref{ic}) is:
\begin{equation}
\begin{array}{cc}
u_t+uu_x+vu_z=- p_x&\\  \delta^2(v_t+uv_x+vv_z)=- p_z&\\
 u_x+v_z=0&\\
 u_z-\delta^2v_x=0&\\
v=\epsilon(\eta_t+u\eta_x) & \textrm{ on }\,
z=1+\epsilon\eta(x,t)\\
 p=\epsilon\eta&
\textrm{ on }\,
z=1+\epsilon\eta(x,t)\\
 v=0 &
\textrm { on } \, z=0.
 \end{array}
\label{e+bc'} \end{equation}
 \\
Making smallness hypotheses on the  parameters $\epsilon$ and $\delta$, one  reduces the  problem to different
physical regimes.
Our analysis is concerned with the shallow-water regime, that is,
\be
\delta<<1.
\ee
 The amplitude of waves is governed by $\epsilon$.
We consider relatively
large amplitude surface waves, meaning that no smallness assumption is made on $\epsilon$.
 For
$\delta=0$, the leading-order system becomes:
\begin{equation}
\begin{array}{cc}
u_t+uu_x+vu_z=- p_x&\\
p_z=0&\\
 u_x+v_z=0&\\
 u_z=0&\\
v=\epsilon(\eta_t+u\eta_x) \,& \textrm{ on }\,
z=1+\epsilon\eta(x,t) \\
p=\epsilon\eta(x,t)\,&  \textrm{ on }\, z=1+\epsilon\eta(x,t)\\
v=0 \,& \textrm { on } z=0.
\end{array}
\label{small}
\end{equation}
The system of equations  (\ref{small}) reduces to
\be
u=u(x,t),\label{13'}
\ee
 \be v=-zu_x, \label{33} \ee
\begin{equation} p=\epsilon\eta(x,t) \label{22}\end{equation}
and
\be\left\{
\begin{array}{ll}
u_t+ u u_x+\epsilon\eta_x=0\\
\epsilon \eta_t+[(1+\epsilon\eta)u]_x=0.
\end{array}
\right.\label{sw1} \ee
 Let us denote by \be H(x,t):=1+\epsilon\eta(x,t).\label{notation} \ee
Then, the system of equations (\ref{sw1}) becomes: \be\left\{
\begin{array}{ll}
u_t+ u u_x+H_x=0\\
H_t+(H u)_x=0,  \end{array} \right. \label{sw2}\ee
that is, the  classical shallow-water equations (see, for
example, \cite{stoker}).
These equations possess an infinite number of integrals of
motion (the conserved quantities) due to Benney \cite{benney} and
can be written in Hamiltonian form relative to a symplectic
structure introduced by Manin \cite{manin}. The second Hamiltonian
structure for the system (\ref{sw2}) was obtained by  Cavalcante
and McKean \cite{caval&mckean}. In fact, the system (\ref{sw2}) is
Hamiltonian with respect to three distinct Hamiltonian structures
\cite{nutku}. These Hamiltonian structures are compatible and
thus, the system of equations (\ref{sw2}) is completely integrable
\cite{olver}. For a rigorous analysis of the system (\ref{sw2}) as
an approximate model of  the water-wave problem see
\cite{lannes1}.

\section{The variational derivation of a new two-component
 shallow-water system}

In what follows we consider $\epsilon$ arbitrary but fixed, there is no smallness assumption on the
 wave amplitude. We are looking for a higher-order correction to the
  classical shallow-water equations (\ref{sw1}),
or (\ref{sw2}) in view of the notation (\ref{notation}). We
observe that the second equation in (\ref{sw2}) is exactly the
second equation of the new two-component
 shallow-water system (\ref{new}).
 The first equation of the  system (\ref{new})
we will derive directly from a variational principle
 in the Lagrangian formalism.

We introduce now the following map
\be
\gamma:\mathbf{R}\times [0,T]\mapsto \mathbf{R}, \quad \gamma(X,t)=x,\label{gamma}
\ee
such that, for a fixed $t$, $\gamma(\cdot, t)$ is an invertible C$^1$-mapping, that is,
\be
\gamma(\cdot, t)\in \textrm{Diff}(\mathbf{R}),
\ee
 and such that
\be
u(x,t)=\gamma_t(X,t),\quad
\textrm{that is},\quad
u(\cdot,t)=\gamma_t\circ \gamma^{-1}\label{u}.
\ee
This map reminds us of the flow map used in the Lagrangian description of the fluid which maps a fluid particle labeled by its initial location $X$ to its later Eulerian position $x$. In the Lagrangian description of the fluid, the Lagrangian velocity $\gamma_t(X,t)$ represents the  velocity of the fluid particle labeled $X$, while the Eulerian velocity $\gamma_t(\gamma^{-1}(x),t)$ represents the velocity of the particle passing the location $x$ at time $t$.\\
In the Eulerian formalism for our problem, for a fixed $t$, $u(x,t)$ can be regarded as a
vector field on $\mathbf{R}$, that is, it belongs to the Lie
algebra of Diff$(\mathbf{R})$. In the Lagrangian formalism for our problem, the velocity phase
 space is the tangent bundle $T\textrm{Diff}(\mathbf{R})$.  For the configuration space Diff($\mathbf{R}$),  we add the technical assumption that the
 smooth functions defined on  $\mathbf{R}$ with value in $\mathbf{R}$
  vanish rapidly at $\pm\infty$ together
 with as many derivatives as necessary.

  The other unknown of our problem is $H(x,t)$, which for a fixed
$t$ can be regarded  as a real function on $\mathbf{R}$,
$H(\cdot,t)\in\mathcal{F}(\mathbf{R})$. We settle that the
evolution equation of $H(x,t)$ is the second equation in
(\ref{sw2}). This equation is an advection equation. In the language of geometry, this equation
 expresses the
fact that  the 1-form
\be
\mathrm{H}(x,t):=H(x,t)dx
\ee
is  Lie transported by the  vector field
\be
\mathrm{u}(x,t):=u(x,t)\pa_x,
\ee
that is, \be \frac{\pa
\mathrm{H}}{\pa t}+\mathrm{L}_{\mathrm{u}}\mathrm{H}=0,\label{lie}
\ee where $\mathrm{L}_{\mathrm{u}}$ denotes the Lie derivative
with respect to the vector field $\mathrm{u}$ (see, for example, \cite{AbMa}
Section 2.2.).
  The
 equation (\ref{lie}) is an
equation written in the Eulerian formalism. With the aid of the
pull back map $\gamma^*$, in the Lagrangian formalism this
becomes: \be \gamma^* \left(\frac{\pa \mathrm{H}}{\pa
t}+\mathrm{L}_{\mathrm{u}}\mathrm{H}\right)=0.\label{6}\ee By
interpreting the Lie derivative of a time-dependent 1-form along a
time-dependent vector field in terms of the flow of the vector
field (see, for example, \cite{AbMa}, Section 2),   we get that\be
\f{d}{dt}\left[\gamma^*(\mathrm{H})\right]=\gamma^*(\mathrm{L}_{\mathrm{u}}\mathrm{H})+
\gamma^*\left(\frac{\pa \mathrm{H}}{\pa
t}\right)\stackrel{(\ref{6})}{=} 0, \ee  that is, we get the
following time invariant 1-form  \be
\mathrm{H}_0:=\gamma^*(\mathrm{H}),\quad
\mathrm{H}_0(X,t)=\mathrm{H}_0(X,0). \ee By the definition of the
pull back map (see, for example, \cite{AbMa}, Section 2), we get
between the components of the 1-forms
$\mathrm{H}_0(X,t):=H_0(X,t)dX$ and $\mathrm{H}(x,t):=H(x,t)dx$
the following relation: \be H_0= (H\circ
\gamma)J_\gamma,\label{h0} \ee where $J_\gamma:=\f{\pa \gamma}{\pa
X}$ is the Jacobian of $\gamma$, or, \be H= (H_0\circ
\gamma^{-1})J_{\gamma^{-1}}.\label{h} \ee

Our goal is to show that the first equation of the system (\ref{new})
yields the critical points of an appropriate action functional
which is completely determined by a scalar function called
Lagrangian. We take the traditional form of the Lagrangian, that
is, the kinetic energy minus the potential energy. In the Eulerian
formalism, taking into account the components (\ref{13'}) and
(\ref{33}) of the velocity field,  the kinetic energy has at the
free surface $z=1+\epsilon\eta(x,t)$ the expression
 \ba E_c(u,\eta)& =& \frac 1{2}
\int_{-\infty}^\infty[u^2+(1+\epsilon\eta)^2u^2_x]dx\nonumber\\
&\stackrel{(\ref{notation})}{=}& \frac 1{2}
\int_{-\infty}^\infty[u^2+H^2u^2_x]dx=:E_c(u,H). \label{energy}\ea
In non-dimensional variables, with $\rho$ and $g$ settled at 1, we
define the gravitational potential energy at the free surface
$z=1+\epsilon\eta(x,t)$, gained by the fluid parcel when it is
vertically displaced from its undisturbed position with $\epsilon
\eta(x,t)$, by \ba E_p(\eta)&=&
\int_{-\infty}^\infty\left(\int_0^{1+\epsilon\eta}(z-1)\,dz\right)dx=\frac
1{2} \int_{-\infty}^\infty(\epsilon\eta)^2dx\nonumber\\
&\stackrel{(\ref{notation})}{=}& \f1{2}\int_{-\infty}^\infty
(H-1)^2dx=:E_p(H). \label{21}\ea  We require in (\ref{energy}) and
(\ref{21}) that at any instant  $t$, \be u\rightarrow 0, \quad
u_x\rightarrow  0\, \textrm{ and }\,     H\rightarrow 1 \textrm{
as } x\rightarrow\pm \infty.\label{limite} \ee Thus, in the
Eulerian formalism,  the Lagrangian function  has the form \be
\mathfrak{L}(u,H)=E_c(u,H)-E_p(H)=
\f1{2}\int_{-\infty}^\infty[u^2+H^2u_x^2-(H-1)^2]dx.\label{lagreuler}
\ee
 Within the Lagrangian formalism, the Lagrangian for our
problem
   will be obtained by transporting the Lagrangian (\ref{lagreuler})
   from the Eulerian formalism,
   to all  tangent spaces
  $T\textrm{Diff}(\mathbf{R})$, this transport being made taking
  into account
(\ref{u}) and (\ref{h}).\\
For each function $H_0\in\mathcal{F}(\mathbf{R})$  independent of
time, we define the Lagrangian
$\mathcal{L}_{H_0}:T\textrm{Diff}(\mathbf{R})\rightarrow\mathbf{R}$
by \ba \mathcal{L}_{H_0}(\gamma,\gamma_t)&:=&
\f1{2}\int_{-\infty}^\infty
\{(\gamma_t\circ\gamma^{-1})^2+[(H_0\circ
\gamma^{-1})J_{\gamma^{-1}}]^2[\partial_x(\gamma_t\circ\gamma^{-1})]^2-\nonumber\\
&& \hspace{2cm} -
[(H_0\circ \gamma^{-1})J_{\gamma^{-1}}-1]^2 \}dx.\label{lagr}\ea
 The Lagrangian $\mathcal{L}_{H_0}$ depends smoothly on $H_0$ and it is right  invariant
under the action of the subgroup \be
\textrm{Diff}(\mathbf{R})_{H_0}=\{\psi\in\textrm{Diff}(\mathbf{R})|
(H_0\circ \psi^{-1})J_{\psi^{-1}}=H_0\},
 \label{subgr}\ee that is, if we
replace the path $\gamma(t,\cdot)$ by
$\gamma(t,\cdot)\circ\psi(\cdot)$, for a fixed time-independent
$\psi$ in Diff($\mathbf{R}$)$_{H_0}$, then $\mathcal{L}_{H_0}$ is
unchanged.\\
The action  on a path $\gamma(t,\cdot)$, $t\in [0,T]$, in
Diff($\mathbf{R}$) is
 \be
 \mathfrak{a}(\gamma):=\int_0^T\mathcal{L}_{H_0}(\gamma,\gamma_t)dt.\label{action''}\ee
The critical points of the action (\ref{action''}) in the space of
paths with fixed endpoints, satisfy \begin{equation}
\f{d}{d\varepsilon} \mathfrak{a}(\gamma+\varepsilon\varphi)\Big
|_{\varepsilon=0}=0,\label{critic}\end{equation} for every path
$\varphi(t,\cdot)$, $t\in[0,T]$, in $\textrm{Diff}(\mathbf{R})$
with endpoints at zero, that is,
\be
\varphi(0,\cdot)=0=\varphi(T,\cdot),\label{endpoints}
\ee and such that
$\gamma+\varepsilon\varphi$ is a small variation of $\gamma$ on
Diff($\mathbf{R}$). With (\ref{lagr}) and
(\ref{action''}) in view, the condition (\ref{critic}) becomes
 \ba \hspace{-0.5cm}
\int_0^T\int_{-\infty}^{\infty}&&\left\{
\left(\gamma_t\circ\gamma^{-1}\right) \f{d}{d\varepsilon}\Big
|_{\varepsilon=0}\left[(\gamma_t+\varepsilon\varphi_t)\circ
(\gamma+\varepsilon\varphi)^{-1}\right
] \right.\nonumber\\
\hspace{-0.5cm} &&
+(H_0\circ\gamma^{-1})J^2_{\gamma^{-1}}\left[\partial_x(\gamma_t\circ\gamma^{-1})\right]^2\f{d}{d\varepsilon}\Big
|_{\varepsilon=0}[H_0\circ(\gamma+\varepsilon\varphi)^{-1}]\nonumber\\
\hspace{-0.5cm} &&
+(H_0\circ\gamma^{-1})^2J_{\gamma^{-1}}\left[\partial_x(\gamma_t\circ\gamma^{-1})\right]^2
\f{d}{d\varepsilon}\Big
|_{\varepsilon=0}[J_{(\gamma+\varepsilon\varphi)^{-1}}]\nonumber\\
\hspace{-0.5cm} && +
[(H_0\circ
\gamma^{-1})J_{\gamma^{-1}}]^2\partial_x(\gamma_t\circ\gamma^{-1})\f{d}{d\varepsilon}\Big
|_{\varepsilon=0}\left[\partial_x\left((\gamma_t+\varepsilon\varphi_t)
\circ(\gamma+\varepsilon\varphi)^{-1}\right)\right]\nonumber\\
\hspace{-0.5cm} &&
-(H_0\circ\gamma^{-1})J^2_{\gamma^{-1}}\f{d}{d\varepsilon}\Big
|_{\varepsilon=0}[H_0\circ(\gamma+\varepsilon\varphi)^{-1}]\nonumber\\
\hspace{-0.5cm} &&
-(H_0\circ\gamma^{-1})^2J_{\gamma^{-1}}\f{d}{d\varepsilon}\Big
|_{\varepsilon=0}[J_{(\gamma+\varepsilon\varphi)^{-1}}]\nonumber\\
\hspace{-0.5cm} && +(J_{\gamma^{-1}})\f{d}{d\varepsilon}\Big
|_{\varepsilon=0}[H_0\circ(\gamma+\varepsilon\varphi)^{-1}]\nonumber\\
\hspace{-0.5cm} &&\left. +
(H_0\circ\gamma^{-1})\f{d}{d\varepsilon}\Big
|_{\varepsilon=0}[J_{(\gamma+\varepsilon\varphi)^{-1}}]
\right\}dxdt=0. \label{66'}\ea
After calculation (for more details
see, for example, \cite{io2012}), we get
\ba
\f{d}{d\varepsilon}\Big
|_{\varepsilon=0}\left[(\gamma_t+\varepsilon\varphi_t)\circ
(\gamma+\varepsilon\varphi)^{-1}\right
]&=&\pa_t(\varphi\circ\gamma^{-1})+(\gamma_t\circ\gamma^{-1})\pa_x(\varphi\circ\gamma^{-1})\nonumber\\
&&-(\varphi\circ\gamma^{-1})\pa_x(\gamma_t\circ\gamma^{-1}),\label{18}
 \ea
 \ba \f{d}{d\varepsilon}\Big
|_{\varepsilon=0}[H_0\circ(\gamma+\varepsilon\varphi)^{-1}]&=&-(\varphi\circ\gamma^{-1})\pa_x(H_0\circ\gamma^{-1})
\label{17} \ea \ba \f{d}{d\varepsilon}\Big
|_{\varepsilon=0}[J_{(\gamma+\varepsilon\varphi)^{-1}}]
&=&-(J_{\gamma^{-1}})\pa_x(\varphi\circ\gamma^{-1})-\pa_x(J_{\gamma^{-1}})
(\varphi\circ\gamma^{-1}),\label{19} \ea
 \begin{eqnarray}  \hspace{-1cm}\f{d}{d\varepsilon}\Big
|_{\varepsilon=0}\left[\partial_x\left((\gamma_t+\varepsilon\varphi_t)
\circ(\gamma+\varepsilon\varphi)^{-1}\right)\right] &=&
\pa_{tx}(\varphi\circ\gamma^{-1})+(\gamma_t\circ\gamma^{-1})\partial^2_x(\varphi\circ\gamma^{-1})\nonumber\\
\hspace{-1cm}&&-[\pa^2_x(\gamma_t\circ\gamma^{-1})](\varphi\circ\gamma^{-1}).
\label{12}\end{eqnarray}
  Thus, from (\ref{18})-(\ref{12}),
the condition (\ref{66'})  becomes:
\begin{eqnarray}
\int_0^T\int_{-\infty}^{\infty}&&
\hspace{-0.3cm}\left\{u\left[\partial_t(\varphi\circ\gamma^{-1})+u\partial_x(\varphi\circ\gamma^{-1})
-(\varphi\circ\gamma^{-1})u_x\right]\right.\nonumber\\
\hspace{1cm}&&
-HH_xu^2_x(\varphi\circ\gamma^{-1})-H^2u^2_x\pa_x(\varphi\circ\gamma^{-1})\nonumber\\
\hspace{1cm}&&+ H^2u_x\left[
\partial_{tx}(\varphi\circ\gamma^{-1})+u\partial^2_x(\varphi\circ\gamma^{-1})
-(\varphi\circ\gamma^{-1})u_{xx}\right]\nonumber \\
\hspace{1cm}&&
+HH_x(\varphi\circ\gamma^{-1})+H^2\pa_x(\varphi\circ\gamma^{-1})\nonumber\\
\hspace{1cm}&& -H_x
(\varphi\circ\gamma^{-1})-H\pa_x(\varphi\circ\gamma^{-1})\left.\right\}dxdt
=0,\label{20}\end{eqnarray}
 where $u=\gamma_t\circ\gamma^{-1}$ and
$H=(H_0\circ\gamma^{-1})J_{\gamma^{-1}}$.
In the above formula, we integrate by parts with respect to
$t$ and $x$, we take into account (\ref{limite}) and (\ref{endpoints}), and
 we get \ba
-\int_0^T\int_{-\infty}^\infty&&\hspace{-0.3cm}(\varphi\circ\gamma^{-1})\left[u_t+3uu_x
-HH_xu^2_x-H^2u_xu_{xx}\right.\nonumber\\
\hspace{1cm}&& -(H^2u_x)_{tx}-(H^2uu_x)_{xx}+HH_x\left.\right]dxdt=0\label{22'} \ea
With $H$ satisfying the second equation
in (\ref{new}), the condition (\ref{22'}) becomes:
\ba
-\int_0^T\int_{-\infty}^\infty&&\hspace{-0.3cm}(\varphi\circ\gamma^{-1})\left\{u_t+3uu_x
+HH_x-\right.\nonumber\\
&&\hspace{2cm}\left.-\left[H^2\left(u_{xt}+uu_{xx}-\f{u^2_x}{2}\right)\right]_x\right\}dxdt=0.\label{23} \ea
Therefore, we proved:

\vspace{0.3cm}

\begin{theorem} For an irrotational
  shallow-water flow,   the non-dimensional horizontal velocity
 of the water $u(x,t)$ and the non-dimensional free upper surface
$H(x,t)=1+\epsilon\eta(x,t)$, for $\epsilon$ arbitrary fixed,
  satisfy the  system (\ref{new}).
  \end{theorem}

\vspace{0.3cm}

We emphasize that for our considerations
we do not require any hypothesis of small amplitude. Under the
additional assumption of a small or moderate amplitude regime,
similar considerations lead to a variational derivation of the
celebrated Korteweg-de Vries  and Camassa-Holm model equations (see \cite{io} and \cite{ckkt}).

\section{The Hamiltonian structure for the  shallow-water system (\ref{new})}

The use of a variational principle in fluid dynamics, beside the aesthetic attraction in condensing the equations by extremizing a scalar quantity,
 retains the Hamiltonian structure with consequent energy conservation. We present below the Hamiltonian structure of the two-component shallow-water
 system (\ref{new}).
\vspace{0.3cm}

\begin{theorem} 
The  shallow-water system (\ref{new}) has the following Hamiltonian form:
\be
\left(\begin{array}{c}
m_t\\
\\
 H_t\end{array}\right)=-\left(\begin{array}{cc}
\pa_x m+m\pa_x & H\pa_x\\
\\
 \pa_x H & 0\end{array}\right)\left(\begin{array}{c}
\f{\delta \mathcal{H}_{N}}{\delta m}\\
\\
 \f{\delta \mathcal{H}_{N}}{\delta H}\end{array}\right),
\ee where $ \mathcal{H}_{N}$ is the total  energy , that is,
\be\mathcal{H}_N(u,H):=E_c(u,H)+E_p(H)=\f
1{2}\int_{-\infty}^\infty\left[
 u^2+H^2u^2_x+(H-1)^2\right]dx,\label{hn}\ee
and $m$ is the momentum density defined by
\be
m:=\f{\delta \mathcal{H}_{N}}{\delta u}=u-(H^2u_x)_x.\label{m}
\ee
\end{theorem}

\vspace{0.3cm}

\begin{proof}
$\f{\delta \mathcal{H}_{N}}{\delta u}$ is the variational derivative of $\mathcal{H}_N$ with respect to $u$, that is,
\be
\f {d}{d\epsilon}\Big
|_{\epsilon=0}\mathcal{H}_N(u+\epsilon \delta u,H)=\int_{-\infty}^\infty \f{\delta \mathcal{H}_{N}}{\delta u} \delta u \, dx. \label{deriv}
\ee
From  the expression (\ref{hn}) of $\mathcal{H}_N$ we have
\be
\f {d}{d\epsilon}\Big
|_{\epsilon=0}\mathcal{H}_N(u+\epsilon \delta u, H)=\int_{-\infty}^\infty\left[
 u\delta u+H^2u_x(\delta u)_x\right]dx.
\ee Integrating by parts and taking  into account (\ref{limite}),
we get \be \f {d}{d\epsilon}\Big
|_{\epsilon=0}\mathcal{H}_N(u+\epsilon \delta u,
H)=\int_{-\infty}^\infty\left[
 u-(H^2u_x)_x\right] \delta u \, dx.
\ee
Therefore, $m$ has the expression (\ref{m}). \\
In order to calculate $\f{\delta \mathcal{H}_{N}}{\delta m}$ and
$\f{\delta \mathcal{H}_{N}}{\delta H}$, that  is, the variational
derivatives of $\mathcal{H}_N$   with respect to $m$ and $H$,
respectively, we write the total energy $\mathcal{H}_{N}$ in terms
of $m$ and $H$. Integrating by parts the second term in the
right-hand integral (\ref{hn}) and taking  into account
(\ref{limite}) we obtain \be \mathcal{H}_N=\f
1{2}\int_{-\infty}^\infty\left[ m u+(H-1)^2\right]dx.\label{hn'}
\ee We can regard (\ref{m}) as an operator equation, that is, \be
m= u-(H^2u_x)_x=:\mathcal{T}_Hu. \label{24}\ee $\mathcal{T}_H$ is
a linear operator defined on the space of real functions $u$
satisfying (\ref{limite}), with the inner product defined by \be
<\mathcal{T}_Hu,v>:=\int_{-\infty}^\infty\left[
uv-(H^2u_x)_xv\right]dx. \label{25}\ee For two functions $u$ and
$v$ satisfying (\ref{limite}), integrating by parts the second
term in the right-hand integral (\ref{25}) we obtain  \be
<\mathcal{T}_Hu,v>=<u,\mathcal{T}_Hv> \ee that is, $\mathcal{T}_H$
is a self-adjoint operator.
 \be
<\mathcal{T}_Hu,u>=\int_{-\infty}^\infty\left[
u^2+H^2u_x^2\right]dx, \ee thus, the operator $\mathcal{T}_H$
 is  positive definite too.\\
The   operator equation (\ref{24}) may  be inverted to determine
$u$  as a continuous function of $m$, \be u=\mathcal{T}_H^{-1}m,
\ee $\mathcal{T}_H^{-1}$ being the inverse operator.\\
 Then, (\ref{hn'}) becomes \be
\mathcal{H}_N(m,H)=\f 1{2}\int_{-\infty}^\infty\left[ m
(\mathcal{T}_H^{-1}m)+(H-1)^2\right]dx.\label{hn''} \ee Let us
calculate now $\f{\delta \mathcal{H}_{N}}{\delta m}$ and
$\f{\delta \mathcal{H}_{N}}{\delta H}$, where \be \f
{d}{d\epsilon}\Big |_{\epsilon=0}\mathcal{H}_N(m+\epsilon \delta
m,H)=\int_{-\infty}^\infty \f{\delta \mathcal{H}_{N}}{\delta m}
\delta m \, dx \label{derivm} \ee and \be \f {d}{d\epsilon}\Big
|_{\epsilon=0}\mathcal{H}_N(m, H+\epsilon \delta
H)=\int_{-\infty}^\infty \f{\delta \mathcal{H}_{N}}{\delta H}
\delta H \, dx. \label{derivH} \ee
 From
(\ref{hn''}), taking into account that $\mathcal{T}_H^{-1}$ is a
linear and self-adjoint operator too, we obtain \ba \f
{d}{d\epsilon}\Big |_{\epsilon=0}\mathcal{H}_N(m+\epsilon \delta
m, H)&=& \f 1{2}\int_{-\infty}^\infty\left[\delta m
(\mathcal{T}_H^{-1}m)+m
(\mathcal{T}_H^{-1}\delta m)\right]dx\nonumber\\
&=& \f 1{2}\int_{-\infty}^\infty\left[\delta m
(\mathcal{T}_H^{-1}m)+
(\mathcal{T}_H^{-1} m)\delta m\right]dx\nonumber\\
&=& \int_{-\infty}^\infty(\mathcal{T}_H^{-1}m)\delta m \, dx. \ea
Therefore, \be \f{\delta \mathcal{H}_{N}}{\delta
m}=\mathcal{T}_H^{-1}m=u. \ee
 From (\ref{hn''}), we also have \ba\f
{d}{d\epsilon}\Big |_{\epsilon=0}\mathcal{H}_N(m, H+\epsilon
\delta  H)&= &\f 1{2}\int_{-\infty}^\infty\left( m \f
d{d\varepsilon}\Big|_{\epsilon=0}\mathcal{T}_{(H+\epsilon
\delta H)}^{-1}m\right)dx \nonumber\\
&&+\int_{-\infty}^\infty(H-1)\delta H\, dx.\label{29} \ea
Differentiating with respect to $\varepsilon$ the identity \be
\mathcal{T}_{(H+\epsilon\delta H)}\circ\mathcal{T}_{(H+\epsilon
\delta H)}^{-1}=Id \ee we get \be \f
d{d\varepsilon}\Big|_{\epsilon=0}\mathcal{T}_{(H+\epsilon \delta
H)}^{-1}m=-\mathcal{T}_{H}^{-1}\left(\f
d{d\varepsilon}\Big|_{\epsilon=0}\mathcal{T}_{(H+\epsilon \delta
H)}u\right).\label{27}\ee Thus, taking into account that
$\mathcal{T}_H^{-1}$ is a self-adjoint operator and the relation
(\ref{27}), the first term in the right-hand side of (\ref{29})
becomes \be \f 1{2}\int_{-\infty}^\infty\left( m \f
d{d\varepsilon}\Big|_{\epsilon=0}\mathcal{T}_{(H+\epsilon \delta
H)}^{-1}m\right)dx =-\f 1{2}\int_{-\infty}^\infty\left(u \f
d{d\varepsilon}\Big|_{\epsilon=0}\mathcal{T}_{(H+\epsilon \delta
H)}u\right)dx.\label{28}
 \ee
From the expression (\ref{24}) of the linear operator
$\mathcal{T}_H$,  \be \f
d{d\varepsilon}\Big|_{\epsilon=0}\mathcal{T}_{(H+\epsilon \delta
H)}u=-2(u_xH_x+Hu_{xx})\delta H-2Hu_x (\delta H)_x.\label{30} \ee
We replace (\ref{30}) in the right-hand side of (\ref{28}), we
integrate by parts, for a function $u$ satisfying (\ref{limite}),
and finally  we obtain \be \f 1{2}\int_{-\infty}^\infty\left( m \f
d{d\varepsilon}\Big|_{\epsilon=0}\mathcal{T}_{(H+\epsilon \delta
H)}^{-1}m\right)dx=
\int_{-\infty}^\infty\left(-Hu^2_x\right)\delta H\, dx.\label{31}
\ee Substituting (\ref{31}) into (\ref{29}) yields
 \be \f {d}{d\epsilon}\Big |_{\epsilon=0}\mathcal{H}_N(m,
H+\epsilon \delta  H)=
\int_{-\infty}^\infty\left(-Hu^2_x+H-1\right)\delta H\, dx, \ee
that is, \be  \f{\delta \mathcal{H}_{N}}{\delta H}=-Hu^2_x+H-1.
\ee It remains to check now that the system \be
\left(\begin{array}{c}
m_t\\
\\
 H_t\end{array}\right)=-\left(\begin{array}{cc}
\pa_x m+m\pa_x & H\pa_x\\
\\
 \pa_x H & 0\end{array}\right)\left(\begin{array}{c}
u\\
-Hu^2_x+H-1\\
\end{array}\right)\label{32}
\ee is the shallow-water system (\ref{new}). It is clear that the
second equation of the system (\ref{32}) is the second equation of
the system (\ref{new}). A straightforward calculation, with $H$
satisfying the second equation in (\ref{new}), shows that the
first equation of the two systems coincide too.\\
 What is left is to show that the operator
 \be
-\left(\begin{array}{cc}
\pa_x m+m\pa_x & H\pa_x\\
\\
 \pa_x H & 0\end{array}\right)
\ee is skew-symmetric and satisfies Jacobi's identity. The verification of Jacobi's identity can be done directly (see, for example, \cite{const97}) or with the assistance of the Lie-Poisson structure (see, for example, \cite{morrison}). This
completes the proof.
\end{proof}

\begin{remark} The Lagrangian (\ref{lagreuler}) does not depend on time and on the space coordinate $x$, that is, it
 is invariant (symmetric) under the time and space translations. Noether's theorem implies
 for each invariance  a unique conservation law (see, for example, \cite{arnold-buch}). Thus, we get for the system of equations
  (\ref{new}) the conservation of the total energy (\ref{hn}) and the conservation of the momentum density (\ref{m}), respectively.
  The local conservation law  for the momentum
  density has the form
  \ba
 m_t&=&-\pa_x(mu)-m\pa_x(u)-H\pa_x\left(-Hu^2_x+H-1\right)\nonumber\\
 &=&-\pa_x\left(mu+\f {u^2}{2}+\f {H^2}{2}+\f{3H^2u^2_x}{2}\right).
 \ea
\end{remark}

\section{Solitary waves for the  shallow-water system (\ref{new})}

We are now interested  in finding the solitary-wave solution of
the nonlinear system (\ref{new}).  For a solution \be
u(x,t)=u(x-ct), \quad H(x,t)=H(x-ct), \ee travelling with speed
$c>0$, the system (\ref{new}) takes the form
\begin{equation}
\left\{\begin{array}{ll}
-cu'+3uu'+ HH'=\left[H^2(u-c)u''-H^2\f{(u')^2}{2}\right]'\\
\\
 (-cH+Hu)'=0.
\end{array}
\right. \label{new1}\end{equation} We require that, at
 any
instant  $t$, \be u\rightarrow 0, \quad u'\rightarrow 0\, \quad
u''\rightarrow 0\, \textrm{ and }\,     H\rightarrow 1 \textrm{ as
} x\rightarrow\pm \infty.\label{limite'} \ee Integrating each
equation of the system (\ref{new1}) and taking into account the
asymptotic limits (\ref{limite'}), we get
\begin{equation}
\left\{\begin{array}{ll}
-cu+\f 3{2} u^2+\f {H^2}{2}=H^2(u-c)u''-H^2\f{(u')^2}{2}+\f 1{2}\\
\\
 u=c\left(1-\f{1}{H}\right).
\end{array}
\right. \label{new2}\end{equation} Plugging the expression of $u$
into the first equation of the system (\ref{new2}) yields an ordinary differential  equation for $H$: \be \f{c^2}{2}-\f {2c^2}{H}+\f
{3c^2}{2} \f 1{H^2}
+\f{H^2}{2}=-c^2\f{H''}{H}+\f{3c^2}{2}\f{(H')^2}{H^2}+\f1{2}.\ee
We multiply the above equation by $2\f {H'}{H^2}$, we integrate,
we take into account the asymptotic limits (\ref{limite'}) and we
obtain \be
-\f{c^2}{H}+\f{2c^2}{H^2}-\f{c^2}{H^3}+H=-c^2\f{(H')^2}{H^3}-\f
1{H}+ 2.\label{34}\ee Now (\ref{34}) becomes \be
c^2(H')^2=(H-1)^2(c^2-H^2). \label{35}\ee From (\ref{35}) it
follows that \be c^2 > H^2, \ee which according to the
asymptotic behavior (\ref{limite'}) of $H$, yields the following
condition for $c$: \be c^2 > 1. \label{36}\ee The solution  of
the separable differential equation (\ref{35}) is obtained by
integration. We denote \be H-1=:\f 1{K}. \label{38}\ee Then, we
get the integral \be I:=\int\f{c\,
dH}{(H-1)\sqrt{c^2-H^2}}=-\int\f {c\,
dK}{\sqrt{(c^2-1)K^2-2K-1}}.\label{37}\ee With the condition
(\ref{36}) in view, we denote \be
\sqrt{(c^2-1)K^2-2K-1}=:w-\sqrt{c^2-1}\,K. \label{38'}\ee In this
way, \be K=\f{w^2+1}{2(w\sqrt{c^2-1}-1)} \ee and the integral
(\ref{37}) becomes \be I=-\int \f{c\, dw}{w\sqrt{c^2-1}-1}=-\f
{c}{\sqrt{c^2-1}}\log(w\sqrt{c^2-1}-1).\ee From the notations
(\ref{38'}) and (\ref{38}), we conclude that \be
I=-\f{c}{\sqrt{c^2-1}}\log\left[\f{\sqrt{c^2-1}\sqrt{c^2-H^2}+c^2-H}{H-1}\right].
\ee Therefore, the solution of the differential equation
(\ref{35}) has the following implicit form \be
\f{\sqrt{c^2-1}\sqrt{c^2-H^2}+c^2-H}{H-1}=\exp\left[-\f{\sqrt{c^2-1}}{c}(x-ct)\right].\label{40}
\ee We add $1$ to both sides of the above equation, we divide by
$\sqrt{c^2-1}$,  we raise to the second power and we get \be
\f{2c^2-H^2-1+2\sqrt{c^2-1}\sqrt{c^2-H^2}}{(H-1)^2}=\left(\f{\exp\left[-\f{\sqrt{c^2-1}}{c}(x-ct)\right]+1}{\sqrt{c^2-1}}\right)^2.
\ee By adding  again $1$ to both sides of the above equation \be
\left(\f{2}{H-1}\right)\f{\sqrt{c^2-1}\sqrt{c^2-H^2}+c^2-H}{H-1}=\left(\f{\exp\left[-\f{\sqrt{c^2-1}}{c}(x-ct)\right]+1}{\sqrt{c^2-1}}\right)^2+1,
\nonumber\ee and by (\ref{40}), we finally obtain \be
\f{2}{H-1}\exp\left[-\f{\sqrt{c^2-1}}{c}(x-ct)\right]=\left(\f{\exp\left[-\f{\sqrt{c^2-1}}{c}(x-ct)\right]+1}{\sqrt{c^2-1}}\right)^2+1.
\ee Thus, we have:

\vspace{0.3cm}
\begin{theorem} The solitary-wave solution of the shallow-water system (\ref{new}) has the form:

\ba
\hspace{-1cm} H(x,t)&=&1+\f{2(c^2-1)\exp\left[\f{\sqrt{c^2-1}}{c}(x-ct)\right]}{c^2\exp\left[2\f{\sqrt{c^2-1}}{c}(x-ct)\right]+
2\exp\left[\f{\sqrt{c^2-1}}{c}(x-ct)\right]+1}\nonumber\\
\hspace{-1cm} & & \hspace{-1.3cm} = 1+\f{c^2-1}{1+\f{c^2+1}{2}\cosh\left[\f{\sqrt{c^2-1}}{c}(x-ct)\right]+\f{c^2-1}{2}\sinh\left[\f{\sqrt{c^2-1}}{c}(x-ct)
\right]}\label{sol1}
\ea
and
\ba
\hspace{-1cm} u(x,t)&=&c\left(1-\f{1}{H(x,t)}\right)
.\label{sol2}
\ea

\end{theorem}






\end{document}